\documentclass{article}%
\usepackage{amsmath}
\usepackage[truetex]{graphicx}%
\usepackage{amsfonts}%
\usepackage{amssymb}
\newtheorem{theorem}{Theorem}

\newtheorem{corollary}[theorem]{Corollary}

\newtheorem{definition}[theorem]{Definition}
\newtheorem{example}[theorem]{Example}

\newenvironment{proof}[1][Proof]{\textbf{#1.} }{\  \rule{0.5em}{0.5em}}
\makeatletter
\@addtoreset{equation}{section}

\makeatother

\begin{document}

\title{Estimation of Complexity for the Ohya-Masuda-Volovich SAT Algorithm}
\author{Shigeo Akashi and Satoshi Iriyama\\Depertment of Information Sciences\\Tokyo University of Science}
\date{}
\maketitle

\begin{abstract}
Ohya and Volovich have been proposed a new quantum computation model with
chaos amplification to solve the SAT problem, which went beyond usual quantum
algorithm. In this paper we study the complexity of the SAT algorithm by
counting the steps of computation algorithm rigorously, which was mentioned in
the paper \cite{OV1,OV2,OV3,OM,AS}. For this purpose, we refine the quantum
gates treating the SAT problem step by step.

\end{abstract}

\section{Introduction}

The problem, asking whether NP-complete problem can be solved in polynomial
time, is one of the most important problems in the computation theory. If the
computational methods are based only on the classical Turing machines, it
seems that the difficulty attached to this problem cannot be removed.

Ohya and Volovich \cite{OV1,OV2,OV3} have been proposed a new quantum
computation algorithm with chaos amplifier to solve the SAT problem\cite{GPFB}%
, which went beyond usual quantum algorithm. This quantum chaos algorithm has
enabled us to solve the SAT problem in a polynomial time\cite{OV1,OV2,OV3,OM},
and moreover alternative solution of the SAT problem is given in the
stochastic limit by Accardi and Ohya \cite{AO}.

In this paper, we study the computational complexity of the SAT algorithm
mentioned in \cite{OV1,OV2,OM,AS} by counting the steps of computation
accurately. For this purpose, we show explicitly how to construct the
Ohya-Masuda algorithm from the elementary gates. In Section 2, the definition
of the SAT problem is explained according to Ohya-Masuda and Accardi-Sabbadini
\cite{OM,AS}. In Section 3, mathematical basis of quantum computation is
given. In Section 4, we determine the number of dust qubits required in the
Ohya-Masuda algorithm exactly, and we construct the unitary operator needed
for computation of the SAT problem. In Section 5, the chaos amplifier
introduced by Ohya-Volovich algorithm is explained. In Section 6, we discuss
the computational complexity of their SAT algorithm.

\section{The SAT Problem}

In this section we review the SAT problem according to Ohya-Masuda,
Accardi-Ohya and Accardi-Sabaddini\cite{OM,AO,AS}. Through this paper,
$\mathbb{N}$ denotes the set of all positive integers, and $\left\{
0,1\right\}  $ denotes the simplest Boolean lattice with the meet-operation
$\wedge$, the join-operation $\vee$ and\ the negation-operation\ $%
\bar{}%
$. Let $n$ be a positive integer and let $X$ be a set consisting of $n$
Boolean variables, which is denoted by $\left\{  x_{1},\dots,x_{n}\right\}  $.
Then, $\bar{X}$ and $X^{\prime}$ denote the two sets consisting of Boolean
variables, which are defined as $\left\{  \bar{x}_{1},\dots,\bar{x}%
_{n}\right\}  $ and $X\cup\bar{X}$, respectively, where $\bar{x}$ means the
negation of $x$. For any subset $C$ of $X^{\prime}$, the truth value of $C$,
which is denoted by $t\left(  C\right)  ,$ is defined as%

\begin{equation}
t\left(  C\right)  \equiv\left\{  \underset{x_{i}\in C\cap X}{\vee}t\left(
x_{i}\right)  \right\}  \vee\left\{  \underset{\bar{x}_{j}\in C\cap\bar{X}%
}{\vee}t\left(  \bar{x}_{j}\right)  \right\}  ,
\end{equation}
where $t\left(  x_{i}\right)  $ and $t\left(  \bar{x}_{j}\right)  $ are the
Boolean values of $x_{i}$ and $\bar{x}_{j}$. $C$ is called a clause, and
$t\left(  C\right)  $ is called the truth value of $C$. Let $m$ be a positive
integer and let $\mathcal{C}$ be a set consisting of $m$ clauses. Then, the
truth value of $\mathcal{C}$, which is denoted by $t\left(  \mathcal{C}%
\right)  $, is defined as%
\begin{equation}
t\left(  \mathcal{C}\right)  =\wedge_{i=1}^{m}t\left(  C_{i}\right)  ,
\end{equation}
where $C_{i}$ is an element of $\mathcal{C}$. It is a matter of course that
the truth value of $\mathcal{C}$ can be exactly determined by $\left\{
\varepsilon_{i}\equiv t\left(  x_{i}\right)  ;i=1,\dots,n\right\}  $.
Therefore, under the above notations, the SAT problem is the problem asking
whether, for a given set of clauses $\mathcal{C}$, there exists an assignment
$\left(  \varepsilon_{1},\dots,\varepsilon_{n}\right)  $ belonging to
$\left\{  0,1\right\}  ^{n}$ and satisfying that $t\left(  \mathcal{C}\right)
=1$ holds. Here, $\mathcal{C}$ is called satisfiable if there exists a
solution $\left(  \varepsilon_{1},\dots,\varepsilon_{n}\right)  $ satisfying
$t\left(  \mathcal{C}\right)  =1$. Here, we can illustrate the following example:

\begin{example}
Let $x_{1},x_{2},x_{3}$ and $x_{4}$ be four Boolean variables, and
$C_{1},C_{2}$ and $C_{3}$ be three clauses defined as $\left\{  x_{1},\bar
{x}_{2},\bar{x}_{3}\right\}  ,\left\{  \bar{x}_{1},x_{2},x_{4}\right\}  $ and
$\left\{  \bar{x}_{1},x_{3},\bar{x}_{4}\right\}  $, respectively. Then,
$\left(  0,0,0,1\right)  $ is a solution of the SAT problem, because, if we
take $0$ as values of $t\left(  x_{1}\right)  ,t\left(  x_{2}\right)  $ and
$t\left(  x_{3}\right)  $, and $1$ as a value of $t\left(  x_{4}\right)  $,
then, the following three equalities:%
\begin{align}
t\left(  C_{1}\right)   &  =t\left(  x_{1}\right)  \vee t\left(  \bar{x}%
_{2}\right)  \vee t\left(  \bar{x}_{3}\right)  =1\nonumber\\
t\left(  C_{2}\right)   &  =t\left(  \bar{x}_{1}\right)  \vee t\left(
x_{2}\right)  \vee t\left(  x_{4}\right)  =1\nonumber\\
t\left(  C_{3}\right)   &  =t\left(  \bar{x}_{1}\right)  \vee t\left(
x_{3}\right)  \vee t\left(  \bar{x}_{4}\right)  =1
\end{align}
hold. Therefore, we can obtain%
\begin{equation}
t\left(  C_{1}\right)  \wedge t\left(  C_{2}\right)  \wedge t\left(
C_{3}\right)  =1.
\end{equation}

These equalities show that $\left\{  C_{1},C_{2},C_{3}\right\}  $ is satisfiable.
\end{example}

\section{Elements of Quantum Computation}

In this section, we review the foundation of quantum computation (see for
instance, \cite{OV3}). Let $\mathbb{C}$ be the set of all complex numbers, and
$\left|  0\right\rangle $ and $\left|  1\right\rangle $ be the two unit
vectors $\binom{1}{0}$ and $\binom{0}{1}$, respectively. Then, for any two
complex numbers $\alpha$ and $\beta$ satisfying $\left|  \alpha\right|
^{2}+\left|  \beta\right|  ^{2}=1$, $\alpha\left|  0\right\rangle
+\beta\left|  1\right\rangle $ is called a qubit. For any positive integer
$N$, let $\mathcal{H}$ be the tensor product Hilbert space defined as $\left(
\mathbb{C}^{2}\right)  ^{\otimes N}$ and let $\left\{  \left|  e_{i}%
\right\rangle ;0\leq i\leq2^{N-1}\right\}  $ be the basis whose elements are
defined as%

\begin{align}
\left|  e_{0}\right\rangle  &  =\left|  0\right\rangle \otimes\left|
0\right\rangle \cdots\otimes\left|  0\right\rangle \equiv\left|
0,0,\cdots,0\right\rangle ,\nonumber\\
\left|  e_{1}\right\rangle  &  =\left|  1\right\rangle \otimes\left|
0\right\rangle \cdots\otimes\left|  0\right\rangle \equiv\left|
1,0,\cdots,0\right\rangle ,\nonumber\\
\left|  e_{2}\right\rangle  &  =\left|  0\right\rangle \otimes\left|
1\right\rangle \cdots\otimes\left|  0\right\rangle \equiv\left|
0,1,\cdots,0\right\rangle ,\nonumber\\
&  \vdots\nonumber\\
\left|  e_{2^{N}-1}\right\rangle  &  =\left|  1\right\rangle \otimes\left|
1\right\rangle \cdots\otimes\left|  1\right\rangle \equiv\left|
1,1,\cdots,1\right\rangle ,
\end{align}
respectively. For any two qubits $\left|  x\right\rangle $ and $\left|
y\right\rangle $, $\left|  x,y\right\rangle $ and $\left|  x^{N}\right\rangle
$ is defined as $\left|  x\right\rangle \otimes\left|  y\right\rangle $ and
$\underset{N\text{ times}}{\underbrace{\left|  x\right\rangle \otimes
\cdots\otimes\left|  x\right\rangle }}$, respectively.

The quantum computation can be formulated mathematically as the multiplication
by unitary operators. Let $U_{NOT}$,$U_{CN}$ and $U_{CCN}$ be the three
unitary operators defined as%
\begin{align}
U_{NOT}  &  \equiv\left|  1\right\rangle \left\langle 0\right|  +\left|
0\right\rangle \left\langle 1\right|  ,\nonumber\\
U_{CN}  &  \equiv\left|  0\right\rangle \left\langle 0\right|  \otimes
I+\left|  1\right\rangle \left\langle 1\right|  \otimes U_{NOT},\nonumber\\
U_{CCN}  &  \equiv\left|  0\right\rangle \left\langle 0\right|  \otimes
I\otimes I+\left|  1\right\rangle \left\langle 1\right|  \otimes\left|
0\right\rangle \left\langle 0\right|  \otimes I+\left|  1\right\rangle
\left\langle 1\right|  \otimes\left|  1\right\rangle \left\langle 1\right|
\otimes U_{NOT}.
\end{align}
$U_{NOT}$,$U_{CN}$ and $U_{CCN}$ is called the NOT-gate, the Controlled-NOT
gate and the Controlled-Controlled-NOT gate, respectively. Moreover, Hadamard
transformation $H$ is defined as the transformation on $\mathbb{C}^{2}$ such
that%
\begin{align}
H\left|  0\right\rangle  &  =\frac{1}{\sqrt{2}}\left(  \left|  0\right\rangle
+\left|  1\right\rangle \right)  ,\nonumber\\
H\left|  1\right\rangle  &  =\frac{1}{\sqrt{2}}\left(  \left|  0\right\rangle
-\left|  1\right\rangle \right)  .
\end{align}

The four operators $U_{NOT}$, $U_{CN}$, $U_{CCN}$ and $H$ are called the
elementary gates. For any $k\in\mathbb{N}$, $U_{H}^{\left(  N\right)  }\left(
k\right)  $ denotes the $k$-tuple Hadamard transformation on $\left(
\mathbb{C}^{2}\right)  ^{\otimes N}$ defined as%

\begin{align}
U_{H}^{\left(  N\right)  }\left(  k\right)  \left|  0^{N}\right\rangle  &
=\frac{1}{2^{k/2}}\left(  \left|  0\right\rangle +\left|  1\right\rangle
\right)  ^{\otimes k}\left|  0^{N-k}\right\rangle \nonumber\\
&  =\frac{1}{2^{k/2}}\sum\limits_{i=0}^{2^{k-1}}\left|  e_{i}\right\rangle
\otimes\left|  0^{N-k}\right\rangle .
\end{align}

These unitary operators can be used for the construction of the following
three unitary operators on $\left(  \mathbb{C}^{2}\right)  ^{\otimes N}$:%

\begin{align}
U_{NOT}^{\left(  N\right)  }\left(  n\right)   &  =I^{\otimes u-1}%
\otimes\left(  \left|  0\right\rangle \left\langle 1\right|  +\left|
1\right\rangle \left\langle 0\right|  \right)  I^{\otimes N-u-1}\\
U_{CN}^{\left(  N\right)  }\left(  u,v\right)   &  =I^{\otimes u-1}%
\otimes\left|  0\right\rangle \left\langle 0\right|  \otimes I^{\otimes
N-u-1}+I^{\otimes u-1}\otimes\left|  1\right\rangle \left\langle 1\right|
\otimes I^{\otimes v-u-1}\otimes U_{NOT}\otimes I^{\otimes N-v-1}\nonumber\\
U_{CCN}^{\left(  N\right)  }\left(  u,v,w\right)   &  =I^{\otimes u-1}%
\otimes\left|  0\right\rangle \left\langle 0\right|  \otimes I^{\otimes
N-u-1}+I^{\otimes u-1}\otimes\left|  1\right\rangle \left\langle 1\right|
\otimes I^{\otimes v-u-1}\otimes\left|  0\right\rangle \left\langle 0\right|
\otimes I^{\otimes N-v-1}\nonumber\\
&  +I^{\otimes u-1}\otimes\left|  1\right\rangle \left\langle 1\right|
\otimes I^{\otimes v-u-1}\otimes\left|  1\right\rangle \left\langle 1\right|
\otimes I^{\otimes w-t-1}\otimes U_{NOT}\otimes I^{\otimes N-w-1},
\end{align}
where $u,v$ and $w$ be a positive integers satisfying $1\leq u<v<w\leq N$.
$U_{NOT}^{\left(  N\right)  }(u)$, $U_{CN}^{\left(  N\right)  }\left(
u,v\right)  $, $U_{CCN}^{\left(  N\right)  }\left(  u,v,w\right)  $ and
$U_{DFT}^{\left(  N\right)  }\left(  k\right)  $ are called $N$-qubit
elementary gates. When no confusion may arise, we identify the $N$-qubit
elementary gates with the elementary gates itself.

Next, three unitary operators $U_{AND},U_{OR}$ and $U_{COPY}$ are called the
logical gates, defined as \cite{AS}%

\begin{align}
U_{AND}  &  \equiv\sum_{\varepsilon_{1},\varepsilon_{2}\in\left\{
0,1\right\}  }\left\{  \left|  \varepsilon_{1},\varepsilon_{2},\varepsilon
_{1}\wedge\varepsilon_{2}\right\rangle \left\langle \varepsilon_{1}%
,\varepsilon_{2},0\right|  +\left|  \varepsilon_{1},\varepsilon_{2}%
,1-\varepsilon_{1}\wedge\varepsilon_{2}\right\rangle \left\langle
\varepsilon_{1},\varepsilon_{2},1\right|  \right\} \nonumber\\
&  =\left|  0,0,0\right\rangle \left\langle 0,0,0\right|  +\left|
0,0,1\right\rangle \left\langle 0,0,1\right|  +\left|  1,0,0\right\rangle
\left\langle 1,0,0\right|  +\left|  1,0,1\right\rangle \left\langle
1,0,1\right| \nonumber\\
&  +\left|  0,1,0\right\rangle \left\langle 0,1,0\right|  +\left|
0,1,1\right\rangle \left\langle 0,1,1\right|  +\left|  1,1,1\right\rangle
\left\langle 1,1,0\right|  +\left|  1,1,0\right\rangle \left\langle
1,1,1\right|  .
\end{align}%

\begin{align}
U_{OR}  &  \equiv\sum_{\varepsilon_{1},\varepsilon_{2}\in\left\{  0,1\right\}
}\left\{  \left|  \varepsilon_{1},\varepsilon_{2},\varepsilon_{1}%
\vee\varepsilon_{2}\right\rangle \left\langle \varepsilon_{1},\varepsilon
_{2},0\right|  +\left|  \varepsilon_{1},\varepsilon_{2},1-\varepsilon_{1}%
\vee\varepsilon_{2}\right\rangle \left\langle \varepsilon_{1},\varepsilon
_{2},1\right|  \right\} \nonumber\\
&  =\left|  0,0,0\right\rangle \left\langle 0,0,0\right|  +\left|
0,0,1\right\rangle \left\langle 0,0,1\right|  +\left|  1,0,1\right\rangle
\left\langle 1,0,0\right|  +\left|  1,0,0\right\rangle \left\langle
1,0,1\right| \nonumber\\
&  +\left|  0,1,1\right\rangle \left\langle 0,1,0\right|  +\left|
0,1,0\right\rangle \left\langle 0,1,1\right|  +\left|  1,1,1\right\rangle
\left\langle 1,1,0\right|  +\left|  1,1,0\right\rangle \left\langle
1,1,1\right|  .
\end{align}%

\begin{align}
U_{COPY}  &  \equiv\sum_{\varepsilon_{1}\in\left\{  0,1\right\}  }\left\{
\left|  \varepsilon_{1},\varepsilon_{1}\right\rangle \left\langle
\varepsilon_{1},0\right|  +\left|  \varepsilon_{1},1-\varepsilon
_{1}\right\rangle \left\langle \varepsilon_{1},1\right|  \right\} \nonumber\\
&  =\left|  0,0\right\rangle \left\langle 0,0\right|  +\left|
0,1\right\rangle \left\langle 0,1\right| \nonumber\\
&  +\left|  1,1\right\rangle \left\langle 1,0\right|  +\left|
1,0\right\rangle \left\langle 1,1\right|  .
\end{align}

$U_{AND},U_{OR}$ and $U_{COPY}$ are called the AND gate, the OR gate and the
COPY gate, respectively. Finally the unitary operators on $\left(
\mathbb{C}^{2}\right)  ^{\otimes N}$, which are denoted by $U_{AND}^{\left(
N\right)  },U_{OR}^{\left(  N\right)  }$ and $U_{COPY}^{\left(  N\right)  }$,
can be defined as%

\begin{align}
U_{AND}^{\left(  N\right)  }(u,v,w)  &  \equiv\sum_{\varepsilon_{1}%
,\varepsilon_{2}\in\left\{  0,1\right\}  }I^{\otimes u-1}\otimes\left|
\varepsilon_{1}\right\rangle \left\langle \varepsilon_{1}\right|  I^{\otimes
v-u-1}\otimes\left|  \varepsilon_{2}\right\rangle \left\langle \varepsilon
_{2}\right| \nonumber\\
&  I^{\otimes w-v-u-1}\otimes\left|  \varepsilon_{1}\wedge\varepsilon
_{2}\right\rangle \left\langle 0\right|  I^{\otimes N-w-v-u}+\nonumber\\
&  I^{\otimes u-1}\otimes\left|  \varepsilon_{1}\right\rangle \left\langle
\varepsilon_{1}\right|  I^{\otimes v-u-1}\otimes\left|  \varepsilon
_{2}\right\rangle \left\langle \varepsilon_{2}\right| \nonumber\\
&  I^{\otimes w-v-u-1}\otimes\left|  1-\varepsilon_{1}\wedge\varepsilon
_{2}\right\rangle \left\langle 1\right|  I^{\otimes N-w-v-u}.
\end{align}%
\begin{align}
U_{OR}^{\left(  N\right)  }\left(  u,v,w\right)   &  \equiv\sum_{\varepsilon
_{1},\varepsilon_{2}\in\left\{  0,1\right\}  }I^{\otimes u-1}\otimes\left|
\varepsilon_{1}\right\rangle \left\langle \varepsilon_{1}\right|  I^{\otimes
v-u-1}\otimes\left|  \varepsilon_{2}\right\rangle \left\langle \varepsilon
_{2}\right| \nonumber\\
&  I^{\otimes w-v-u-1}\otimes\left|  \varepsilon_{1}\vee\varepsilon
_{2}\right\rangle \left\langle 0\right|  I^{\otimes N-w-v-u}+\nonumber\\
&  I^{\otimes u-1}\otimes\left|  \varepsilon_{1}\right\rangle \left\langle
\varepsilon_{1}\right|  I^{\otimes v-u-1}\otimes\left|  \varepsilon
_{2}\right\rangle \left\langle \varepsilon_{2}\right| \nonumber\\
&  I^{\otimes w-v-u-1}\otimes\left|  1-\varepsilon_{1}\vee\varepsilon
_{2}\right\rangle \left\langle 1\right|  I^{\otimes N-w-v-u}.
\end{align}%
\begin{align}
U_{COPY}^{\left(  N\right)  }\left(  u,v\right)   &  \equiv\sum_{\varepsilon
_{1}\in\left\{  0,1\right\}  }I^{\otimes u-1}\left|  \varepsilon
_{1}\right\rangle \left\langle \varepsilon_{1}\right|  I^{\otimes
v-u-1}\left|  \varepsilon_{1}\right\rangle \left\langle 0\right|  I^{\otimes
N-v-u}\nonumber\\
&  +I^{\otimes u-1}\left|  \varepsilon_{1}\right\rangle \left\langle
\varepsilon_{1}\right|  I^{\otimes v-u-1}\left|  1-\varepsilon_{1}%
\right\rangle \left\langle 1\right|  I^{\otimes N-v-u}.
\end{align}
where $u,v$ and $w$ are positive integers satisfying $1\leq u<v<w\leq N$.
These operators can be represented, in terms of elementary gates, as%
\begin{align}
U_{OR}^{\left(  N\right)  }\left(  u,v,w\right)   &  =U_{CN}^{\left(
N\right)  }\left(  u,w\right)  \cdot U_{CN}^{\left(  N\right)  }\left(
v,w\right)  \cdot U_{CCN}^{\left(  N\right)  }\left(  u,v,w\right)
,\nonumber\\
U_{AND}^{\left(  N\right)  }\left(  u,v,w\right)   &  =U_{CCN}^{\left(
N\right)  }\left(  u,v,w\right)  ,\nonumber\\
U_{COPY}^{\left(  N\right)  }\left(  u,v\right)   &  =U_{CN}^{\left(
N\right)  }\left(  u,v\right)  .
\end{align}

\section{Quantum Computational Model of the Ohya-Masuda Algorithm}

In this section, we explain the computation method which has been developed by
Ohya-Masuda and Accardi-Sabbadini\cite{OM,AS}. The quantum algorithm is
described by a combination of the unitary operators on a Hilbert space
$\mathcal{H}$. Throughout this section, let $n$ be the total number of Boolean
variables used in the SAT problem. Let $\mathcal{C}$ be a set of clauses whose
cardinality is equal to $m$. Following the method of the Ohya-Masuda
algorithm\cite{OM}, let $\mathcal{H}=\left(  \mathbf{C}^{2}\right)  ^{\otimes
n+\mu+1}$ be a Hilbert space and $\left|  v_{in}\right\rangle $ be the initial
state $\left|  v_{in}\right\rangle =\left|  0^{n},0^{\mu},0\right\rangle $,
where $\mu$ is the number of dust qubits which is determined by the following
theorem 2. Let $U_{\mathcal{C}}^{\left(  n\right)  }$ be a unitary operator
satisfying the following equation.%

\begin{align}
U_{\mathcal{C}}^{\left(  n\right)  }\left|  v_{in}\right\rangle  &
=\frac{1}{\sqrt{2^{n}}}\sum_{i=0}^{2^{n}-1}\left|  e_{i},x^{\mu},t_{e_{i}%
}\left(  \mathcal{C}\right)  \right\rangle \nonumber\\
&  \equiv\left|  v_{out}\right\rangle
\end{align}
where $x^{\mu}$ denotes a $\mu$ strings of binary symbols and $t_{e_{i}%
}\left(  \mathcal{C}\right)  $ is a truth value of $\mathcal{C}$ with $e_{i}$.
In \cite{AS}, a method to construct $U_{\mathcal{C}}^{\left(  n\right)  }$ is
discussed. Let $\left\{  s_{k};k=1,\dots,m\right\}  $ be the sequence defined
as%
\begin{align}
s_{1}  &  =n+1,\nonumber\\
s_{2}  &  =s_{1}+card\left(  C_{1}\right)  +\delta_{1,card\left(
C_{1}\right)  }-1,\nonumber\\
s_{i}  &  =s_{i-1}+card\left(  C_{i-1}\right)  +\delta_{1,card\left(
C_{i-1}\right)  },\text{ \ \ }3\leq i\leq m,
\end{align}
where $card\left(  C_{i}\right)  $ means the cardinality of a clause $C_{i}$.
And let $s_{f}$ be a number defined as%
\begin{equation}
s_{f}=s_{m}-1+card\left(  C_{m}\right)  +\delta_{1,card\left(  C_{m}\right)
}.
\end{equation}

Then we can prove the following:

\begin{theorem}
For $m\geq2$, the total number of dust qubits $\mu$ is
\begin{align}
\mu &  =s_{f}-1-n\nonumber\\
&  =\sum_{k=1}^{m}card\left(  C_{k}\right)  +\delta_{1,card\left(
C_{k}\right)  }-2.
\end{align}
\end{theorem}

\begin{proof}
If $card\left(  C_{k}\right)  $ is greater than 1, it is required to use the
join operation $\left(  card\left(  C_{k}\right)  -1\right)  $ times to obtain
the value of $t\left(  C_{k}\right)  $. If $card\left(  C_{k}\right)  $ is
equal to $1$. we prepare one qubit to make a copy of a Boolean variable
included in $C_{k}$. Here, assume that there exists a qubit where
$\wedge_{i=1}^{k-1}t\left(  C_{i}\right)  $ is stored. Then, one more qubit is
required to store $\left(  \wedge_{i=1}^{k-1}t\left(  C_{i}\right)  \right)
\wedge t\left(  C_{k}\right)  $. These results imply that $card\left(
C_{k}\right)  -1+\delta_{1,card\left(  C_{k}\right)  }+1$ qubits are required
to compute $t\left(  C_{k}\right)  $ and $\wedge_{i=1}^{k-1}t\left(
C_{i}\right)  $. Therefore, we can obtain%
\begin{equation}
s_{k+1}-s_{k}=card\left(  C_{k}\right)  +\delta_{1,card\left(  C_{k}\right)
}.
\end{equation}

Finally, the total number of dust qubits $\mu$ which are required to compute
$\wedge_{i=1}^{m}t\left(  C_{i}\right)  $ is%
\begin{align}
\mu &  =\sum_{k=1}^{m}s_{k+1}-s_{k}\nonumber\\
&  =s_{2}-s_{1}+\sum_{k=2}^{m}card\left(  C_{k}\right)  +\delta_{1,card\left(
C_{k}\right)  }\nonumber\\
&  =\sum_{k=1}^{m}card\left(  C_{k}\right)  +\delta_{1,card\left(
C_{k}\right)  }-2\nonumber\\
&  =s_{f}-1-n
\end{align}
\end{proof}

Determining $\mu$ and the work spaces for computing $t\left(  C_{k}\right)  $,
we can construct $U_{\mathcal{C}}^{\left(  n\right)  }$ concretely.

\begin{theorem}
\label{5}The unitary operator $U_{\mathcal{C}}^{\left(  n\right)  }$, is
represented as%
\begin{equation}
U_{\mathcal{C}}^{\left(  n\right)  }=\prod_{i=m-1}^{1}U_{AND}^{\left(
n+\mu+1\right)  }\left(  i\right)  \prod_{j=m}^{1}U_{OR}^{\left(
n+\mu+1\right)  }\left(  j\right)  U_{H}^{\left(  n+\mu+1\right)  }\left(
n\right)  .
\end{equation}
\end{theorem}

\begin{proof}
For any positive integers $u,v$ and $w$ satisfying $1\leq u<v<w\leq n+\mu+1$,
if $C_{k}\cap\left\{  x_{u},\bar{x}_{u}\right\}  \neq\phi$ and $C_{k}%
\cap\left\{  x_{v},\bar{x}_{v}\right\}  \neq\phi$ hold, $U_{OR}^{\left(
N\right)  }\left(  u,v,w\right)  $\ is defined as%
\begin{equation}
U_{OR}^{\left(  N\right)  }\left(  u,v,w\right)  =\left\{
\begin{array}
[c]{c}%
U_{NOT}^{\left(  N\right)  }\left(  u\right)  U_{OR}^{\left(  N\right)
}\left(  u,v,w\right)  U_{NOT}^{\left(  N\right)  }\left(  u\right)  ,\text{
}\bar{x}_{u},x_{v}\in C_{k},\\
U_{NOT}^{\left(  N\right)  }\left(  v\right)  U_{OR}^{\left(  N\right)
}\left(  u,v,w\right)  U_{NOT}^{\left(  N\right)  }\left(  v\right)  ,\text{
}x_{u},\bar{x}_{v}\in C_{k},\\
U_{NOT}^{\left(  N\right)  }\left(  u\right)  U_{NOT}^{\left(  N\right)
}\left(  v\right)  U_{OR}^{\left(  N\right)  }\left(  u,v,w\right)
U_{NOT}^{\left(  N\right)  }\left(  v\right)  U_{NOT}^{\left(  N\right)
}\left(  u\right)  ,\text{ }\bar{x}_{u},\bar{x}_{v}\in C_{k}.
\end{array}
\right.
\end{equation}

If the cardinality of $C_{k}$ is equal to one, then there exists a Boolean
variable $x_{u}$ satisfying $C_{k}=\left\{  x_{u}\right\}  $ or $C_{k}%
=\left\{  \bar{x}_{u}\right\}  $. Therefore, $U_{OR}^{\left(  N\right)
}\left(  k\right)  $ is defined as%
\begin{equation}
U_{OR}^{\left(  N\right)  }\left(  k\right)  =\left\{
\begin{array}
[c]{c}%
U_{COPY}^{\left(  N\right)  }\left(  u,s_{k}\right)  ,\text{ }x_{u}\in
C_{k},\\
U_{NOT}^{\left(  N\right)  }\left(  s_{k}\right)  U_{COPY}^{\left(  N\right)
}\left(  u,s_{k}\right)  ,\text{ }\bar{x}_{u}\in C_{k}.
\end{array}
\right.
\end{equation}

If the cardinality of $C_{k}$ is equal to two, then there exists two Boolean
variables $x_{u}$ and $x_{v}$ satisfying that either $x_{u}\in C_{k}$ or
$\bar{x}_{u}\in C_{k}$ holds, and moreover, either $x_{v}\in C_{k}$ or
$\bar{x}_{v}\in C_{k}$ holds. Therefore, $U_{OR}^{\left(  N\right)  }\left(
k\right)  $ is defined as
\begin{equation}
U_{OR}^{\left(  N\right)  }\left(  k\right)  =U_{OR}^{\left(  N\right)
}\left(  u,v,s_{k}\right)  .
\end{equation}

If the cardinality of $C_{k}$ is greater than 2, $U_{OR}^{\left(  N\right)
}\left(  k\right)  $ can be defined by the way as above, namely, this operator
is defined as
\begin{equation}
U_{OR}^{\left(  N\right)  }\left(  k\right)  =\prod_{i=card\left(
C_{k}\right)  -2}^{1}U_{OR}^{\left(  N\right)  }\left(  w,s_{k}+i-1,s_{k}%
+i\right)  \cdot U_{OR}^{\left(  N\right)  }\left(  u,v,s_{k}\right)  .
\end{equation}

If the cardinality of $\mathcal{C}$ is equal to one, then $U_{AND}^{\left(
N\right)  }\left(  1\right)  $ is defined as%
\begin{equation}
U_{AND}^{\left(  N\right)  }\left(  1\right)  =U_{COPY}^{\left(  N\right)
}\left(  s_{1}+card\left(  C_{1}\right)  +\delta_{1,card\left(  C_{1}\right)
}-1,s_{1}+card\left(  C_{1}\right)  +\delta_{1,card\left(  C_{1}\right)
}\right)  .
\end{equation}

If the cardinality of $\mathcal{C}$ is greater than one, then $U_{AND}%
^{\left(  N\right)  }\left(  k\right)  $ is defined as
\begin{equation}
U_{AND}^{\left(  N\right)  }\left(  k\right)  \left\{
\begin{array}
[c]{c}%
=U_{AND}^{\left(  N\right)  }\left(  s_{k+1}-1,s_{k+2}-2,s_{k+2}-1\right)
,\text{ \ \ }1\leq k\leq m-2,\\
=U_{AND}^{\left(  N\right)  }\left(  s_{m}-1,s_{m}+card\left(  C_{m}\right)
+\delta_{1,card\left(  C_{m}\right)  }-2,\right.  \text{\ \ \ \ \ }\\
\text{ \ \ \ \ \ }\left.  s_{m}+card\left(  C_{m}\right)  +\delta
_{1,card\left(  C_{m}\right)  }-1\right)  ,\text{ }\ \ k=m-1.
\end{array}
\right.
\end{equation}

It is clear that $U_{AND}^{\left(  N\right)  }\left(  k\right)  $ can compute
$\wedge_{i=1}^{k-1}t\left(  C_{i}\right)  $. We can construct the unitary
operator $U_{\mathcal{C}}^{\left(  n\right)  }$ from $\left\{  U_{OR}^{\left(
n+\mu+1\right)  }\left(  i\right)  ;1\leq i\leq m\right\}  $ and $\left\{
U_{AND}^{\left(  n+\mu+1\right)  }\left(  i\right)  ;1\leq i\leq m-1\right\}
$ as follows:
\begin{equation}
U_{\mathcal{C}}^{\left(  n\right)  }=\prod_{i=m-1}^{1}U_{AND}^{\left(
n+\mu+1\right)  }\left(  i\right)  \prod_{j=m}^{1}U_{OR}^{\left(
n+\mu+1\right)  }\left(  j\right)  U_{H}^{\left(  n+\mu+1\right)  }\left(
n\right)  .
\end{equation}
\end{proof}

The following theorem is shown in Accardi-Ohya \cite{AO}.

\begin{theorem}
$\mathcal{C}$ is SAT if and only if%
\begin{equation}
P_{n+\mu,1}U_{\mathcal{C}}^{\left(  n\right)  }\left|  v_{in}\right\rangle
\neq0
\end{equation}
where $P_{n+\mu,1}$ denotes the projector%
\begin{equation}
P_{n+\mu,1}\equiv I^{\otimes n+\mu}\otimes\left|  1\right\rangle \left\langle
1\right|
\end{equation}
onto the subspace of $\mathcal{H}$ spanned by the vectors $\left|
\varepsilon^{n},\varepsilon^{\mu},1\right\rangle $.
\end{theorem}

\subsection{Example}

For example, Let $x_{1},x_{2},x_{3}$ and $x_{4}$ be four Boolean variables,
and $C_{1},C_{2},C_{3}$ and $C_{4}$ be four clauses defined as $\left\{
x_{1},x_{4},\bar{x}_{2}\right\}  ,\left\{  x_{2},x_{3,}x_{4}\right\}
,\left\{  x_{1},\bar{x}_{3}\right\}  $ and $\left\{  x_{3},\bar{x}_{1},\bar
{x}_{2}\right\}  $, respectively. Let $\mathcal{C}$ be the set of clauses
consisting of $C_{1},C_{2},C_{3}$ and $C_{4}.$First, we calculate $s_{1}%
,s_{2},s_{3},s_{4}$ and $s_{f}$. According to Theorem \ref{11}, we obtain%

\begin{align}
s_{1}  &  =n+1=5,\nonumber\\
s_{2}  &  =s_{1}+card\left(  C_{1}\right)  +\delta_{1,card\left(
C_{1}\right)  }-1\nonumber\\
&  =5+3+0-1\nonumber\\
&  =7,\nonumber\\
s_{3}  &  =s_{2}+card\left(  C_{2}\right)  +\delta_{1,card\left(
C_{2}\right)  }\nonumber\\
&  =7+3+0\nonumber\\
&  =10,\nonumber\\
s_{4}  &  =s_{3}+card\left(  C_{3}\right)  +\delta_{1,card\left(
C_{3}\right)  }\nonumber\\
&  =10+2\nonumber\\
&  =12,\nonumber\\
s_{f}  &  =s_{4}+card\left(  C_{4}\right)  +\delta_{1,card\left(
C_{4}\right)  }-1\nonumber\\
&  =12+3-1\nonumber\\
&  =14.
\end{align}

Then we construct OR and AND gates following Theorem \ref{5}. We have%

\begin{align}
U_{OR}^{\left(  14\right)  }\left(  1\right)   &  =U_{NOT}^{\left(  14\right)
}(2)U_{OR}^{\left(  14\right)  }\left(  2,5,6\right)  U_{NOT}^{\left(
14\right)  }(2)U_{OR}^{\left(  14\right)  }\left(  1,4,5\right)  ,\nonumber\\
U_{OR}^{\left(  14\right)  }\left(  2\right)   &  =U_{OR}\left(  4,7,8\right)
U_{OR}\left(  2,3,7\right)  ,\nonumber\\
U_{OR}^{\left(  14\right)  }\left(  3\right)   &  =U_{NOT}^{\left(  14\right)
}(3)U_{OR}^{\left(  14\right)  }\left(  1,3,10\right)  U_{NOT}^{\left(
14\right)  }(3),\nonumber\\
U_{OR}^{\left(  14\right)  }\left(  4\right)   &  =U_{NOT}^{\left(  14\right)
}(2)U_{OR}^{\left(  14\right)  }\left(  2,12,13\right)  U_{NOT}^{\left(
14\right)  }(2)U_{NOT}^{\left(  14\right)  }(1)U_{OR}^{\left(  14\right)
}\left(  3,1,12\right)  U_{NOT}^{\left(  14\right)  }(1),
\end{align}%

\begin{align}
U_{AND}^{\left(  14\right)  }\left(  1\right)   &  =U_{AND}^{\left(
14\right)  }\left(  6,8,9\right)  ,\nonumber\\
U_{AND}^{\left(  14\right)  }\left(  2\right)   &  =U_{AND}^{\left(
14\right)  }\left(  9,10,11\right)  ,\nonumber\\
U_{AND}^{\left(  14\right)  }\left(  3\right)   &  =U_{AND}^{\left(
14\right)  }\left(  11,13,14\right)  .
\end{align}

Thus, we obtain the unitary gate $U_{\mathcal{C}}^{\left(  4\right)  }$ by the
combination of the above gates as%

\begin{align}
U_{\mathcal{C}}^{\left(  4\right)  }  &  =U_{AND}^{\left(  14\right)  }\left(
11,13,14\right)  U_{AND}^{\left(  14\right)  }\left(  9,10,11\right)
U_{AND}^{\left(  14\right)  }\left(  6,8,9\right) \nonumber\\
&  \cdot U_{NOT}^{\left(  14\right)  }(2)U_{OR}^{\left(  14\right)  }\left(
2,12,13\right)  U_{NOT}^{\left(  14\right)  }(2)U_{NOT}^{\left(  14\right)
}(1)U_{OR}^{\left(  14\right)  }\left(  3,1,12\right)  U_{NOT}^{\left(
14\right)  }(1)\nonumber\\
&  \cdot U_{NOT}^{\left(  14\right)  }(3)U_{OR}^{\left(  14\right)  }\left(
1,3,10\right)  U_{NOT}^{\left(  14\right)  }(3)\nonumber\\
&  \cdot U_{OR}^{\left(  14\right)  }\left(  4,7,8\right)  U_{OR}^{\left(
14\right)  }\left(  2,3,7\right) \nonumber\\
&  \cdot U_{NOT}^{\left(  14\right)  }(2)U_{OR}^{\left(  14\right)  }\left(
2,5,6\right)  U_{NOT}^{\left(  14\right)  }(2)U_{OR}^{\left(  14\right)
}\left(  1,4,5\right)  .
\end{align}

Let $\left|  v_{in}\right\rangle $ be the initial state $\left|
v_{in}\right\rangle $ $=\left|  0^{4},0^{10},0\right\rangle $. Applying
$U_{DFT}^{\left(  14\right)  }\left(  4\right)  $ to $\left|  v_{in}%
\right\rangle $, we have%

\begin{align}
\left|  v_{in}\right\rangle  &  \equiv U_{DFT}^{\left(  14\right)  }\left(
4\right)  \left|  0^{4},0^{9},0\right\rangle \nonumber\\
&  =\frac{1}{\left(  \sqrt{2}\right)  ^{4}}\sum_{i=0}^{2^{4}-1}\left|
e_{i},0^{9},0\right\rangle \nonumber\\
&  =\frac{1}{\left(  \sqrt{2}\right)  ^{4}}\sum_{\varepsilon_{1}%
,\varepsilon_{2},\varepsilon_{3},\varepsilon_{4}\in\left\{  0,1\right\}
}\left|  \varepsilon_{1},\varepsilon_{2},\varepsilon_{3},\varepsilon_{4}%
,0^{9},0\right\rangle \nonumber\\
&  \equiv\left|  v\right\rangle .
\end{align}

Next, applying $%
{\textstyle\prod_{k=4}^{1}}
U_{OR}^{\left(  14\right)  }\left(  k\right)  $ to $\left|  v\right\rangle $,
we obtain%

\begin{align}
&  U_{OR}^{\left(  14\right)  }\left(  4\right)  U_{OR}^{\left(  14\right)
}\left(  3\right)  U_{OR}^{\left(  14\right)  }\left(  2\right)
U_{OR}^{\left(  14\right)  }\left(  1\right)  \left|  v\right\rangle
\nonumber\\
&  =\frac{1}{\left(  \sqrt{2}\right)  ^{4}}U_{OR}^{\left(  14\right)  }\left(
4\right)  U_{OR}^{\left(  14\right)  }\left(  3\right)  U_{OR}^{\left(
14\right)  }\left(  2\right)  U_{OR}^{\left(  14\right)  }\left(  1\right)
\sum_{\varepsilon_{1},\varepsilon_{2},\varepsilon_{3},\varepsilon_{4}%
\in\left\{  0,1\right\}  }\left|  \varepsilon_{1},\varepsilon_{2}%
,\varepsilon_{3},\varepsilon_{4},0^{9},0\right\rangle \nonumber\\
&  =\frac{1}{\left(  \sqrt{2}\right)  ^{4}}U_{OR}^{\left(  14\right)  }\left(
4\right)  U_{OR}^{\left(  14\right)  }\left(  3\right)  U_{OR}^{\left(
14\right)  }\left(  2\right)  \sum_{\varepsilon_{1},\varepsilon_{2}%
,\varepsilon_{3},\varepsilon_{4}\in\left\{  0,1\right\}  }\left|
\varepsilon_{1},\varepsilon_{2},\varepsilon_{3},\varepsilon_{4},\varepsilon
_{1}\vee\varepsilon_{4},\varepsilon_{1}\vee\varepsilon_{4}\vee\bar
{\varepsilon}_{2},0^{8},0\right\rangle \nonumber\\
&  =\frac{1}{\left(  \sqrt{2}\right)  ^{4}}\sum_{\varepsilon_{1}%
,\varepsilon_{2},\varepsilon_{3},\varepsilon_{4}\in\left\{  0,1\right\}
}\left|  \varepsilon_{1},\varepsilon_{2},\varepsilon_{3},\varepsilon
_{4},\varepsilon_{1}\vee\varepsilon_{4},\varepsilon_{1}\vee\varepsilon_{4}%
\vee\overline{\varepsilon_{2}},\varepsilon_{2}\vee\varepsilon_{3}%
,\varepsilon_{2}\vee\varepsilon_{3}\vee\varepsilon_{4},0,\right. \nonumber\\
&  \left.  \varepsilon_{1}\vee\bar{\varepsilon}_{3},0,\varepsilon_{3}\vee
\bar{\varepsilon}_{1},\varepsilon_{3}\vee\bar{\varepsilon}_{1}\vee
\bar{\varepsilon}_{2},0\right\rangle \nonumber\\
&  \equiv\left|  v^{\prime}\right\rangle .
\end{align}

Finally, applying AND gates to $\left|  v^{\prime}\right\rangle $, we have%

\begin{align}
U_{AND}^{\left(  14\right)  }\left(  m\right)  \left|  v^{\prime
}\right\rangle  &  =U_{AND}^{\left(  14\right)  }\left(  11,14,15\right)
U_{AND}^{\left(  14\right)  }\left(  9,10,11\right)  U_{AND}^{\left(
14\right)  }\left(  6,8,9\right)  \left|  v^{\prime}\right\rangle \nonumber\\
&  =\frac{1}{\left(  \sqrt{2}\right)  ^{4}}\sum_{\varepsilon_{1}%
,\varepsilon_{2},\varepsilon_{3},\varepsilon_{4}\in\left\{  0,1\right\}
}\left|  \varepsilon_{1},\varepsilon_{2},\varepsilon_{3},\varepsilon
_{4},\varepsilon_{1}\vee\varepsilon_{4},\varepsilon_{1}\vee\varepsilon_{4}%
\vee\overline{\varepsilon_{2}},\varepsilon_{2}\vee\varepsilon_{3}%
,\varepsilon_{2}\vee\varepsilon_{3}\vee\varepsilon_{4},\right. \nonumber\\
&  \left.  t\left(  C_{1}\right)  \wedge t\left(  C_{2}\right)  ,\varepsilon
_{1}\vee\bar{\varepsilon}_{3},t\left(  C_{1}\right)  \wedge t\left(
C_{2}\right)  \wedge t\left(  C_{3}\right)  ,\varepsilon_{3}\vee
\bar{\varepsilon}_{1},\varepsilon_{3}\vee\bar{\varepsilon}_{1}\vee
\bar{\varepsilon}_{2},t_{\varepsilon}\left(  \mathcal{C}\right)  \right\rangle
.
\end{align}

After the measurement of the last qubit, we obtain the final state%

\begin{equation}
\rho^{\prime}=\frac{7}{16}\left|  1\right\rangle \left\langle 1\right|
+\frac{9}{16}\left|  0\right\rangle \left\langle 0\right|  .
\end{equation}

\section{Chaos Amplification of the SAT algorithm}

Let us explain the chaos amplifier introduced by Ohya-Volovich \cite{OV1,OV2}.
Let $T\left(  \mathcal{C}\right)  $ be the set of all the elements in
$\left\{  0,1\right\}  ^{n}$ satisfying $t\left(  \mathcal{C}\right)  =1$ and
$\left|  T\left(  \mathcal{C}\right)  \right|  $ be the cardinality of
$T\left(  \mathcal{C}\right)  $. After the quantum computation due to the
Ohya-Masuda algorithm, the quantum computer will be in the state%
\begin{equation}
\left|  v_{out}\right\rangle =\sqrt{1-q^{2}}\left|  \varphi_{0}\right\rangle
\otimes\left|  0\right\rangle +q\left|  \varphi_{1}\right\rangle
\otimes\left|  1\right\rangle ,
\end{equation}
where $\left|  \varphi_{0}\right\rangle $ and $\left|  \varphi_{1}%
\right\rangle $ are normalized $n$ qubit states and $q=\sqrt{\left|  T\left(
\mathcal{C}\right)  \right|  /2^{n}}$. It is useful to quantum computing in
which the result probability of unitary computation is very small. Let $E_{0}$
and $E_{1}$ be a projection operators $E_{0}=\left|  0\right\rangle
\left\langle 0\right|  $ and $E_{1}=\left|  1\right\rangle \left\langle
1\right|  $. According to the Ohya-Volovich algorithm\cite{OV1,OV2}, we
transform the state $\left|  v_{out}\right\rangle $ into the density matrix of
the form%

\begin{equation}
\rho=q^{2}E_{1}+\left(  1-q^{2}\right)  E_{0}.
\end{equation}

The logistic map which is given by the equation%
\begin{equation}
x_{n+1}=ax_{n}\left(  1-x_{n}\right)  \equiv f_{a}\left(  x_{n}\right)
,\text{ \ }x_{n}\in\left[  0,1\right]  .
\end{equation}
The properties of this map depend on the parameter $a$. Then the density
matrix $\rho$ above is interpreted as the initial data $\rho_{0}$, and
Ohya-Volovich applied the logistic map to the state $\rho$ as
\begin{equation}
\rho_{m}=\frac{(I+f_{a}^{m}(\rho_{0})\sigma_{3})}{2}, \label{12}%
\end{equation}
where $I$ is the identity matrix and $\sigma_{3}$ is the $z$-component of
Pauli matrix on $\mathbb{C}^{2}.$ Finally the value of $\sigma_{3}$ is
measured in the state $\rho_{m}$%

\begin{equation}
M_{m}\equiv tr\rho_{m}\sigma_{3}.
\end{equation}

The following theorems \ref{8}, \ref{2} and \ref{9} are proven in
\cite{OV1,OV2}.

\begin{theorem}
\label{8}%
\begin{equation}
\rho_{m}=\frac{(I+f_{a}^{m}(q^{2})\sigma_{3})}{2},\text{ and }M_{m}=f_{a}%
^{m}(q^{2}).
\end{equation}
\end{theorem}

\begin{theorem}
\label{2}For the logistic map $x_{n+1}=ax_{n}\left(  1-x_{n}\right)  $ with
$a\in\left[  0,4\right]  $ and $x_{0}\in\left[  0,1\right]  $, let $x_{0}$ be
$\frac{1}{2^{n}}$ and a set $J$ be $\left\{  0,1,2,\dots,n,\dots,2n\right\}
$. If $a$ is $3.71$, then there exists an integer $m$ in $J$ satisfying
$x_{m}>\frac{1}{2}.$
\end{theorem}

\begin{theorem}
\label{9}Let $a$ and $n$ be the same in above proposition. If there exists $m
$ in $J$ such that $x_{m}>\frac{1}{2},$ then $m>\frac{n-1}{\log_{2}3.71}.$
\end{theorem}

From these theorems, we have

\begin{corollary}
Let $\rho$ be the initial state of the Ohya-Masuda algorithm and $\rho_{0}$ be
the initial data of the chaos amplifire correspond to $\rho$. For all $m$, it
holds%
\begin{equation}
M_{m}\left\{
\begin{array}
[c]{c}%
=0\text{ \ \ \ iff }\mathcal{C}\text{ is not SAT}\\
>0\text{ \ \ \ iff }\mathcal{C}\text{ is SAT \ \ \ \ }%
\end{array}
\right.
\end{equation}
\end{corollary}

\begin{corollary}
\label{7}Let $x_{0}\equiv\frac{r}{2^{n}}$ with $r\equiv\left|  T\left(
\mathcal{C}\right)  \right|  $. From Theorem \ref{9}, there exists $m$
satisfying the following inequation if $\mathcal{C}$ is SAT.
\begin{equation}
\left[  \frac{n-1-\log_{2}r}{\log_{2}3.71-1}\right]  \leq m\leq\left[
\frac{5}{4}\left(  n-1\right)  \right]  .
\end{equation}
\end{corollary}

\begin{figure}
[ptb]
\begin{center}
\includegraphics[
height=3.0943in,
width=4.8144in
]%
{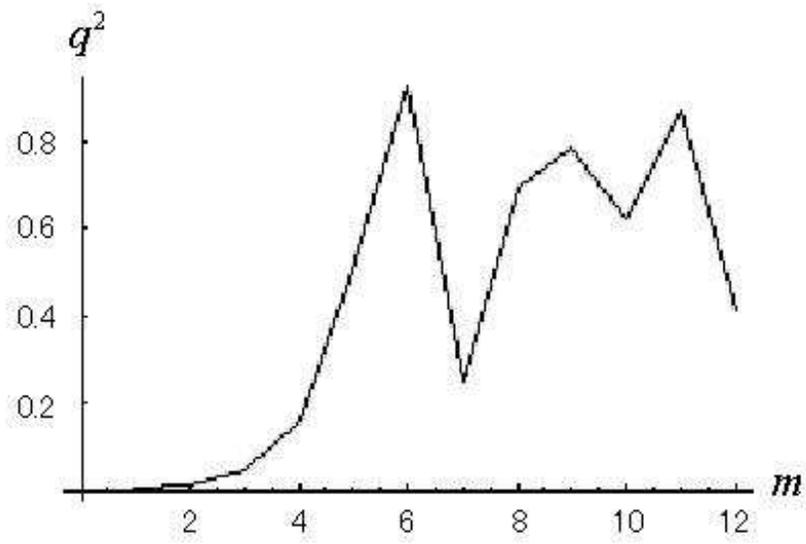}%
\caption{Amplification process, $n=12$ and $l=1.$}%
\label{1}%
\end{center}
\end{figure}

We plot $q^{2}$ as a function of $m$ with $n=12$ and $l=1$. In Figure \ref{1},
it is shown that $q^{2}$ increases in the first six steps. We can see that
$q^{2}$ becomes more than $\frac{1}{2}$ within $\frac{n-1}{\log_{2}3.71-1}$ steps.%

\begin{figure}
[ptb]
\begin{center}
\includegraphics[
height=3.0009in,
width=4.8144in
]%
{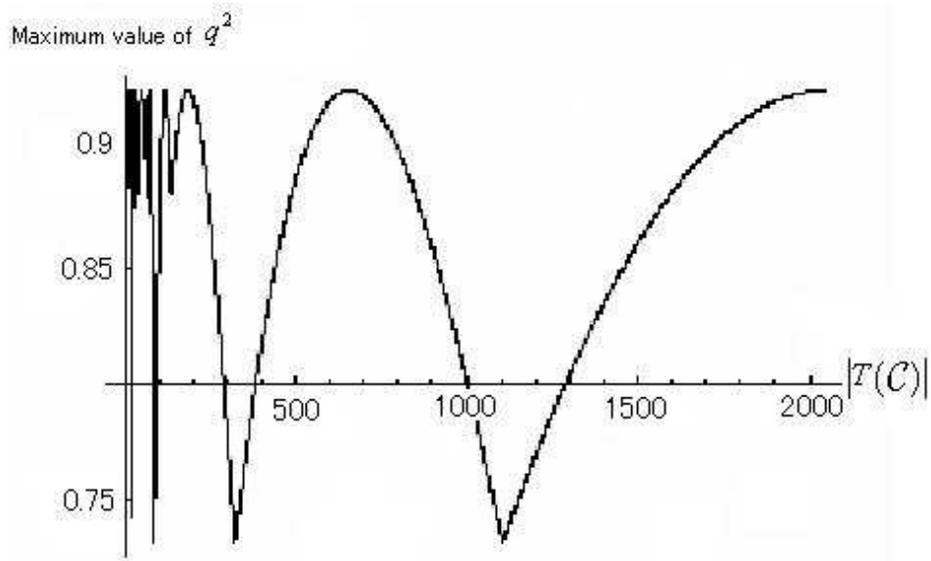}%
\caption{Maximum value of $q^{2}$}%
\label{10}%
\end{center}
\end{figure}

If we apply the chaos amplifier to $\rho_{0}$ $\frac{n-1-\log_{2}r}{\log
_{2}3.71-1}$ times, then the maximum value of $q^{2}$ can be achieved as we
can see Figure 2. In other words, Figure \ref{10} shows that, $\frac{n-1-\log
_{2}r}{\log_{2}3.71-1}$ times application of the chaos amplifier can obtain
the maximum value of $q^{2}$ which is greater than or equal to $\frac{1}{2}$.%

\section{\bigskip Computational Complexity of the OMV SAT Algorithm}

In this section, we define the computational complexity of the OMV SAT
algorithm including the chaos amplifier. First, we define the computational
complexity of the quantum part, Ohya-Masuda algorithm. The computational
complexity of a quantum algorithm is determined by the number of elementary
gates in the algorithm. Since the unitary operator $U_{\mathcal{C}}^{\left(
n\right)  }$ which has been used in the Ohya-Masuda algorithm is constructed
by the product of elementary gates (see Theorem \ref{5}), we define the
computational complexity of $U_{\mathcal{C}}^{\left(  n\right)  }$ as the
number of the elementary gates.

\begin{definition}
The computational complexity of the unitary operator $U$ consisting of the
elementary gates, denoted by $T_{Q}\left(  U\right)  $, is defined as%
\begin{equation}
T_{Q}\left(  U\right)  =\left|  U\right|  ,
\end{equation}
where $\left|  U\right|  $ denotes the number of elementary gates which are
the components of $U$.
\end{definition}

Next, we define the computational complexity of the chaos amplifier as follows.

\begin{definition}
For any positive integer $n$, we define the computational complexity of the
chaos amplifier, denoted by $T_{C}\left(  n\right)  $, is defined as%
\begin{equation}
T_{C}\left(  n\right)  =\max\left\{  m;m=\min\left\{  l;M_{l}\left(
q^{2}\right)  \geq\frac{1}{2}\right\}  ,q^{2}\in\left\{  \frac{1}{2^{n}%
},\frac{2}{2^{n}},\cdots,\frac{2^{n-1}}{2^{n}}\right\}  \right\}  .
\end{equation}
\end{definition}

\begin{corollary}
The computational complexity of the chaos amplifier of the SAT algorithm with
$n$ Boolean variables can be obtained as
\begin{equation}
T_{C}\left(  n\right)  =\left[  \frac{5}{4}\left(  n-1\right)  \right]  .
\label{3}%
\end{equation}
\end{corollary}

In the SAT algorithm with $n$ Boolean variables, according to Corollary
\ref{7}, there exists a proper $m$ satisfying $M_{m}\geq\frac{1}{2}$ within
$m\leq\left[  \frac{5}{4}\left(  n-1\right)  \right]  $. Since it is
impossible to know the value of $m$ satisfying $M_{m}\geq\frac{1}{2}$ before
the computation of the chaos amplifier, we have to compute again when we chose
$m $ not satisfying $M_{m}\geq\frac{1}{2}$. It implies that we must repeat the
quantum computation $\frac{5}{4}\left(  n-1\right)  $ times at worst. Thus, we
define the computational complexity of the SAT algorithm as the product of
$T_{Q}\left(  U_{\mathcal{C}}^{\left(  n\right)  }\right)  $ and $T_{C}\left(
n\right)  $.

\begin{definition}
The computational complexity of the SAT algorithm is defined as%
\begin{equation}
T_{Q}\left(  U_{\mathcal{C}}^{\left(  n\right)  }\right)  T_{C}\left(
n\right)  .
\end{equation}
\end{definition}

\begin{theorem}
For a set of clauses $\mathcal{C}$ and $n$ Boolean variables, the
computational complexity of the SAT algorithm including the chaos amplifier,
denoted by $T\left(  \mathcal{C},n\right)  $, is obtained as follows.%
\begin{align}
T\left(  \mathcal{C},n\right)   &  =T_{Q}\left(  U_{\mathcal{C}}^{\left(
n\right)  }\right)  T_{C}\left(  n\right) \nonumber\\
&  =\{3\sum_{k=1}^{m}\left(  card\left(  C_{k}\right)  -1\right)  +\sum
_{k-1}^{m}2card\left(  C_{k}\cap\left\{  \bar{x}_{1},\dots,\bar{x}%
_{n}\right\}  \right) \nonumber\\
&  +m-1+n\}\left[  \frac{5}{4}\left(  n-1\right)  \right] \nonumber\\
&  \leq\left(  8mn-2m-1\right)  \left[  \frac{5}{4}\left(  n-1\right)  \right]
\nonumber\\
&  =\mathcal{O}\left(  poly\left(  n\right)  \right)  , \label{4}%
\end{align}
where $poly\left(  n\right)  $ denotes a polynomial of $n$.
\end{theorem}

\begin{proof}
Since $T_{Q}\left(  \prod_{k=m}^{1}U_{OR}^{\left(  n+\mu+1\right)  }\left(
k\right)  \right)  ,T_{Q}\left(  \prod_{k=m-1}^{1}U_{AND}^{\left(
n+\mu+1\right)  }\left(  k\right)  \right)  $ and $T_{Q}\left(  U_{DFT}%
^{\left(  n+\mu+1\right)  }\left(  n\right)  \right)  $ can be estimated by
the next inequalities:%
\begin{align}
T_{Q}\left(  \prod_{k=m}^{1}U_{OR}^{\left(  n+\mu+1\right)  }\left(  k\right)
\right)   &  =3\sum_{k=1}^{m}\left(  card\left(  C_{k}\right)  -1\right)
+\sum_{k-1}^{m}2card\left(  C_{k}\cap\left\{  \bar{x}_{1},\dots,\bar{x}%
_{n}\right\}  \right) \nonumber\\
&  \leq3m\left(  2n-1\right)  +2mn,\nonumber\\
T_{Q}\left(  \prod_{k=m-1}^{1}U_{AND}^{\left(  n+\mu+1\right)  }\left(
k\right)  \right)   &  =m-1,\nonumber\\
T_{Q}\left(  U_{DFT}^{\left(  n+\mu+1\right)  }\left(  n\right)  \right)   &
=n. \label{6}%
\end{align}

Then, $T_{Q}\left(  U_{\mathcal{C}}^{\left(  n\right)  }\right)  $ can be
obtained by the next inequalities:%
\begin{align}
T_{Q}\left(  U_{\mathcal{C}}^{\left(  n\right)  }\right)   &  =T_{Q}\left(
\prod_{k=m}^{1}U_{OR}^{\left(  n+\mu+1\right)  }\left(  k\right)  \right)
+T_{Q}\left(  \prod_{k=m-1}^{1}U_{AND}^{\left(  n+\mu+1\right)  }\left(
k\right)  \right) \nonumber\\
&  +T_{Q}\left(  U_{DFT}^{\left(  n+\mu+1\right)  }\left(  n\right)  \right)
\nonumber\\
&  =3\sum_{k=1}^{m}\left(  card\left(  C_{k}\right)  -1\right)  +\sum
_{k-1}^{m}2card\left(  C_{k}\cap\left\{  \bar{x}_{1},\dots,\bar{x}%
_{n}\right\}  \right) \nonumber\\
&  +m-1+n\nonumber\\
&  \leq8mn-2m+n-1. \label{11}%
\end{align}

Since $T_{C}\left(  n\right)  =\left[  \frac{5}{4}\left(  n-1\right)  \right]
$, we obtain (\ref{4}).
\end{proof}

\section{Conclusion}

In this paper, we have determined the number of dust qubits $\mu$ exactly and
constructed $U_{\mathcal{C}}^{\left(  n\right)  }$ step by step. Moreover, we
set the computational complexity of the Ohya-Masuda algorithm and
Ohya-Volovich algorithm, and computed $T\left(  \mathcal{C},n\right)  $
accurately. Therefore, the Ohya-Volovich algorithm can give us a practical
method to solve the SAT problem.


\begin{thebibliography}{9}                                                                                                %

\bibitem {OV1}M.Ohya and I.V.Volovich, \textit{Quantum computing and chaotic
amplification}, J. opt. B, Quantum Computing, 2003.

\bibitem {OV2}M.Ohya and I.V.Volovich, \textit{New quantum algorithm for
studying NP-complete problems}, Rep. on Math. Phys. 2003.

\bibitem {OV3}M.Ohya and I.V.Volovich, \textit{Quantum information,
computation, cryptography and teleportation}, Springer (to appear).

\bibitem {GPFB}J.Gu, P.W.Purdom, J.Franco, and B.W.Wah, ''Algorithms for the
Satisfiability (SAT) Problem: a Survey,''\ Preliminary version, 1996. http://citeseer.nj.nec.com/56722.html

\bibitem {OM}M.Ohya and N.Masuda, \textit{NP problem in Quantum Algorithm,}
Open Systems and Information Dynamics, 7 No.1 (2000), 33-39.

\bibitem {AO}L.Accardi and M.Ohya, \textit{A stochastic limit approach to the
SAT problem}, Proc. of SCI2003.

\bibitem {AS}L.Accardi and R.Sabbadini, On the Ohya--Masuda quantum SAT
Algorithm, Preprint Volterra, N. 432, 2000.
\end{thebibliography}
\end{document}